\documentclass{elsart}
\usepackage{times}
\usepackage{epsfig}
\usepackage{marvosym}
\usepackage{amsmath}
\usepackage{amsfonts}
\usepackage{amssymb}
\begin{document}
\bibliographystyle{roman}

\journal{Nuclear Instruments and Methods A}

\def\hb{\hfill\break}
\def\MeV{\rm MeV}
\def\GeV{\rm GeV}
\def\TeV{\rm TeV}

\def\m{\rm m}
\def\cm{\rm cm}
\def\mm{\rm mm}
\def\lam{$\lambda_{\rm int}$}
\def\rad{$X_0$}
 
\def\NIM{Nucl. Instr. and Meth.~}
\def\ieee {{IEEE Trans. Nucl. Sci.~}}

\def\etal{{\it et al.}}
\def\eg{{\it e.g.,~}}
\def\ie{{\it i.e.,~}}
\def\cf{{\it cf.~}}
\def\etc{{\it etc.~}}
\def\vs{{\it vs.~}}
\begin{frontmatter}
\title{Dual-Readout Calorimetry with Lead Tungstate Crystals}

\author{N. Akchurin$^a$, L. Berntzon$^a$, A. Cardini$^b$, R. Ferrari$^c$, G. Gaudio$^c$,}
\author{J. Hauptman$^d$, H. Kim$^a$, L. La Rotonda$^e$, M. Livan$^c$, E. Meoni$^e$,}
\author{H. Paar$^f$, A. Penzo$^g$, D. Pinci$^h$, A. Policicchio$^e$,} 
\author{S. Popescu$^{i,}$\thanksref{Leave}}
\author{G.~Susinno$^e$, Y. Roh$^a$, W. Vandelli$^c$ and R. Wigmans$^{a,}$\thanksref{Corres}}

\address{$^a$ Texas Tech University, Lubbock (TX), USA\\
$^b$ Dipartimento di Fisica, Universit\`a di Cagliari and INFN Sezione di Cagliari, Italy\\
$^c$ Dipartimento di Fisica Nucleare e Teorica, Universit\`a di Pavia and INFN Sezione di Pavia, Italy\\
$^d$ Iowa State University, Ames (IA), USA\\
$^e$ Dipartimento di Fisica, Universit\'a della Calabria and INFN Cosenza, Italy\\
$^f$ University of California at San Diego, La Jolla (CA), USA\\
$^g$ INFN Trieste, Italy\\
$^h$ Dipartimento di Fisica, Universit\`a di Roma ''La Sapienza''  and INFN Sezione di Roma\\
$^i$ CERN, Gen\`eve, Switzerland}
\thanks[Leave]{On leave from IFIN-HH, Bucharest, Romania.}
\thanks[Corres]{Corresponding author.
              Email wigmans@ttu.edu, fax (+1) 806 742-1182.}

\begin{abstract}
Results are presented of beam tests in which a small electromagnetic calorimeter consisting of lead tungstate crystals was exposed to 50 GeV electrons and pions. This calorimeter was backed up by the DREAM Dual-Readout calorimeter, which measures the scintillation and \v{C}erenkov light produced in the shower development, using two different media. The signals from the crystal calorimeter were analyzed in great detail in an attempt to determine the contributions from these two types of light to the signals, event by event. This information makes it possible to eliminate the dominating source of fluctuations and thus achieve an important improvement in hadronic calorimeter performance.
\end{abstract}

\end{frontmatter}

\section{Introduction}

High-precision measurements of hadrons and hadron jets have become increasingly important in 
experimental particle physics. Such measurements are considered a crucial ingredient of
experiments at a future high-energy Linear Electron-Positron Collider.
Historically, by far the best performance in this respect has been delivered by {\sl compensating}
hadron calorimeters \cite{RWbook}. In these instruments, the response to the electromagnetic (em) and non-electromagnetic shower components is equalized by design, and therefore the detrimental effects
of event-to-event fluctuations in the energy sharing between these components are eliminated.
These effects include hadronic signal non-linearity, a poor hadronic energy resolution, especially at high energies where deviations from $E^{-1/2}$ scaling become the dominant factor, and a non-Gaussian response function. 

In recent years, an alternative technique has been developed: The Dual Readout Method (DREAM).  DREAM calorimeters offer the same advantages as compensating ones, and are not subject to the
disadvantages of the latter. These disadvantages derive from the fact that compensation can only be achieved in sampling calorimeters with a small sampling fraction (\eg 2.3\% in lead/plastic-scintillator), plus the fact that compensation relies upon efficient detection of the neutrons produced in the shower development. These requirements limit the electromagnetic (em) energy resolution achievable with these instruments, while the excellent hadronic performance is only achieved with sufficiently large detector volumes and integration times. 

DREAM calorimeters are based on a simultaneous measurement of different types of signals which provide complementary information about details of the shower development. The first calorimeter of this type that we developed and tested (in the context of a generic R\&D project) was based on a copper absorber structure, equipped with two types of active media. Scintillating fibers measure the total energy deposited
by the shower particles, while \v{C}erenkov light is only produced by the charged, relativistic shower particles.
Since the latter are almost exclusively found in the em shower component (dominated by
$\pi^0$s produced in hadronic showers), a comparison of the two signals makes it possible to measure the 
energy fraction carried by this component, $f_{\rm em}$, event by event. As a result, the effects of fluctuations
in this component, which are  responsible for all traditional problems in non-compensating calorimeters, can be eliminated. This leads to an important improvement in the
hadronic calorimeter performance.
The performance characteristics of this detector are described elsewhere \cite{DREAMem,DREAMhad,DREAMmu}. 

Once the effects of the dominant source of fluctuations, \ie fluctuations in the em energy fraction $f_{\rm em}$, are eliminated, the performance characteristics are determined (and limited) by other types of fluctuations.
In the described detector, a prominent role was played by the small number of \v{C}erenkov photoelectrons constituting the signals (8 p.e./GeV). However, there is absolutely no reason why the
DREAM principle would only work in fiber calorimeters, or even in sampling calorimeters for that matter.
One could in principle even use a homogeneous (fully sensitive) detector, provided that the light signals can be separated into scintillation and \v{C}erenkov components.
In this paper, we describe the results of tests of this idea. A small electromagnetic calorimeter made of
lead tungstate (PbWO$_4$) crystals was tested in conjunction with the DREAM calorimeter mentioned above, and exposed to high-energy particle beams at CERN's Super Proton Synchrotron.

In Section 2, we describe the detectors and the
experimental setup in which they were tested. In Section 3, we discuss the experimental
data that were taken and the methods used to analyze these data. 
In Section 4, the experimental results are described and discussed. A summary and conclusions are presented in Section 5.    

\section{Detectors and experimental setup}

\subsection{Detectors}

\begin{figure}[htb]
\epsfysize=6cm
\centerline{\epsffile{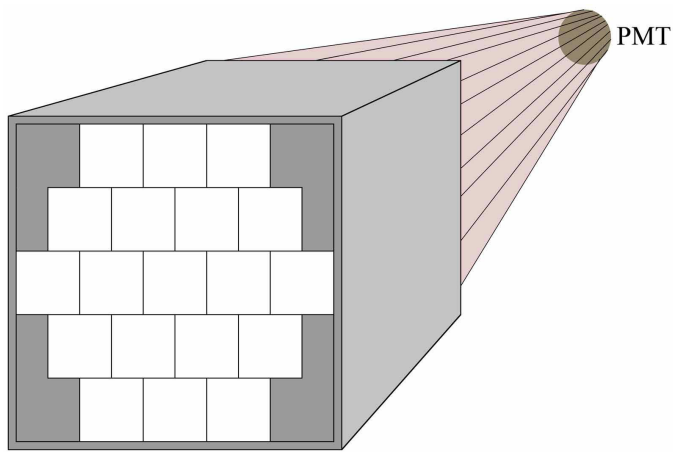}}
\caption{\small
The lead tungstate electromagnetic section of the calorimeter system.}
\label{ECAL}
\end{figure}

The calorimeter system used in these experiments comprised two sections.The electromagnetic section (ECAL) consisted of 19 lead tungstate (PbWO$_4$) crystals\footnote{On loan from the ALICE Collaboration, who use these crystals for their PHOS calorimeter.}. Each crystal was
18 cm long, with a cross section of $2.2\times 2.2$ cm$^2$. These crystals were arranged in a matrix, as shown in Figure \ref{ECAL}. 

For the purpose of these tests, this ensemble of crystals was considered one unit. The crystals were not optically isolated from each other, and the light produced by showering particles was read out by only two photomultiplier tubes (PMT)\footnote{Hamamatsu R5900U, 10-stage.}, one located at each end of the crystal matrix. The light was funneled into these tubes by means of 
cones made of  aluminized mylar, a highly reflective material.

For the hadronic section (HCAL) of the calorimeter system, we used the original DREAM calorimeter \cite{DREAMem,DREAMhad,DREAMmu}. 
The basic element of this detector is an extruded copper rod, 2 meters long
and 4$\times$ 4 mm$^2$ in cross section. This rod is hollow, and the central cylinder has a diameter of 2.5 mm. 
Seven optical fibers were inserted in this hole. Three of these were plastic scintillating fibers, the other four fibers were undoped fibers, intended for detecting \v{C}erenkov light.
The instrumented volume had a
length of 2.0 m (10 $\lambda_{\rm int}$, 100 $X_0$), an effective radius of 16.2 cm and a mass of 1030 kg. 

The fibers were grouped to form 19 hexagonal towers. The effective radius of each tower was 37.1 mm ($1.82 \rho_M$).
A central tower (\#1) was surrounded by two hexagonal rings, the Inner Ring (6 towers, numbered 2-7) and the Outer Ring (12 towers,
numbered 8-19). The towers were longitudinally unsegmented. 
The fibers sticking out at the rear end of this structure
were separated into 38 bunches: 19 bunches of scintillating fibers and 19 bunches of \v{C}erenkov fibers. In this way, the
readout structure was established. Each bunch was coupled through a 2 mm air gap to a 
PMT\footnote{Hamamatsu R-580, 10-stage, 1.5 inch diameter, bialkali photocathode, borosilicate window.}.
More information about this detector is given elsewhere \cite{DREAMem,DREAMhad}.

\subsection{The beam line}

The measurements described in this paper were performed in the H4 beam line of the Super Proton Synchrotron
at CERN. The detectors
were mounted on a platform that could move vertically and sideways with respect to the beam.
The ECAL was rotated into different positions (see Section 4) by hand.
Two small scintillation counters provided the signals that were used to trigger the data acquisition system.
These Trigger Counters were 2.5 mm thick, and the area of overlap was 6$\times$6 cm$^2$. A coincidence between the logic signals from these counters provided the trigger.

\subsection{Data acquisition}

Measurement of the time structure of the calorimeter signals formed a very important part of the tests
described here. In order to limit distortion of this structure as much as possible, we used 15 mm thick air-core cables to transport the ECAL signals to the counting room. Such cables were also used for
the signals from the trigger counters, and these were routed such as to
minimize delays in the DAQ system\footnote{We measured the signal speed to be 0.78$c$ in these cables.}.

The HCAL signals  were transported through RG-58 cables with (for timing purposes) appropriate lengths to the counting room.
The ECAL signals  were split (passively) into 5 equal parts in the counting room. One part was sent to a charge ADC, the other 4 signals were used for analysis of the time structure by means of a FADC. The latter unit measured the amplitude of the signals at a rate of 200 MHz.
During a time interval of 80 ns, 16 measurements of the amplitude were thus obtained. The 4 signals from the splitter box were measured separately in 4 different channels of the FADC module\footnote{Dr. Struck SIS3320,  http://www.struck.de/sis3320.htm}. Signals 2, 3 and 
4 were delayed by 1.25 ns, 2.50 ns and 3.75 ns with respect to signal 1. By using 4 channels of the FADC module in this way, the time structure of the signals was thus effectively measured with a resolution of 1.25 ns (800 MHz).
\begin{figure}[htb]
\epsfysize=8cm
\centerline{\epsffile{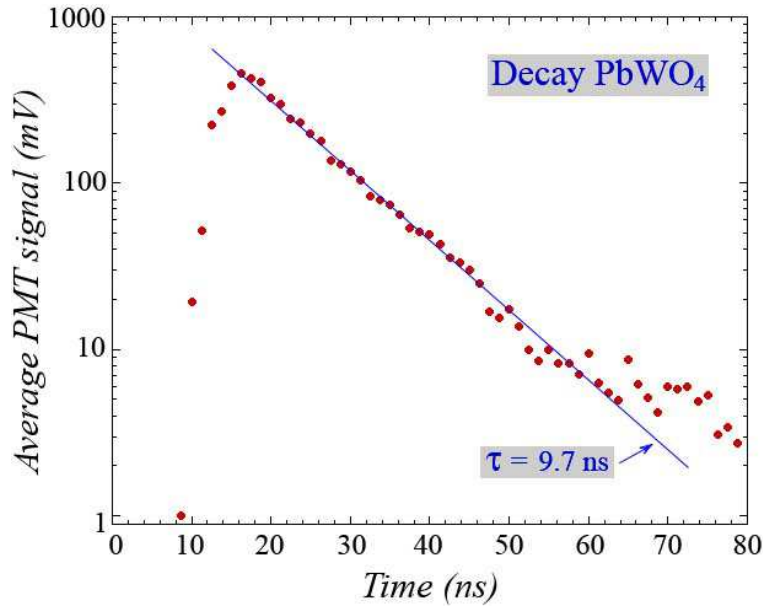}}
\caption{\small
Average time structure of the signals from 50 GeV electron showers in the lead tungstate crystals.}
\label{esignal}
\end{figure}

The quality of the information obtained in this way is illustrated in Figure \ref{esignal}, which shows the 
average time structure of the signals from 50 GeV electron showers developing in the lead tungstate ECAL. The trailing edge of these signals is well described by an exponential decay with a time constant of 9.7 ns.

The charge measurements were performed with 12-bit LeCroy 1182 ADCs.
These had a sensitivity of 50 fC/count and a conversion time of 16 $\mu$s.
The ADC gate width was 100 ns, and the calorimeter signals arrived $\sim 20$ ns after the start of the gate.

The data acquisition system used VME electronics.
A single VME crate hosted all the needed readout and control boards.
The trigger logic was implemented through NIM modules and the signals were sent 
to a VME I/O register, which also collected the spill and the global 
busy information. The VME crate was linked to a Linux based computer 
through an SBS 620\footnote{http://www.gefanucembedded.com/products/457} 
optical VME-PCI interface that allowed memory 
mapping of the VME resources via an open source driver\footnote{http://www.awa.tohoku.ac.jp/$\sim$sanshiro/kinoko-e/vmedrv/}.
The computer was equipped with a 2 GHz Pentium-4 
CPU, 1 GB of RAM, and was running a CERN SLC 4.3 operating system\footnote{http://linux.web.cern.ch/linux/scientific4/}.

The data acquisition was based on a single-event polling mechanism and 
performed by a pair of independent programs that communicated
through a first-in-first-out buffer, built on top of a 32 MB shared 
memory. Only exclusive accesses were allowed and concurrent requests were 
synchronised with semaphores. The chosen scheme 
optimized the CPU utilization and increased the data taking efficiency by 
exploiting the bunch structure of the SPS, where beam particles were provided to
our experiment during a spill of 4.8 s, out of a total cycle time of 16.8 s.
During the spill, the readout program collected data from the VME modules and 
stored them into the shared memory, with small access times. During the remainder of the SPS cycle, a 
recorder program dumped the events to the disk. Moreover, the buffer
presence allowed low-priority monitoring programs to run (off-spill) in 
spy mode. With this scheme, we were able to reach a data acquisition rate 
as high as 2 kHz, limited by the FADC readout time. 
The typical event size was $\sim 1$ kB.  
All calorimeter signals and the signals from the auxiliary detectors were monitored on-line.

\subsection{Calibration of the detectors}

\begin{figure}[htb]
\epsfysize=5.5cm
\centerline{\epsffile{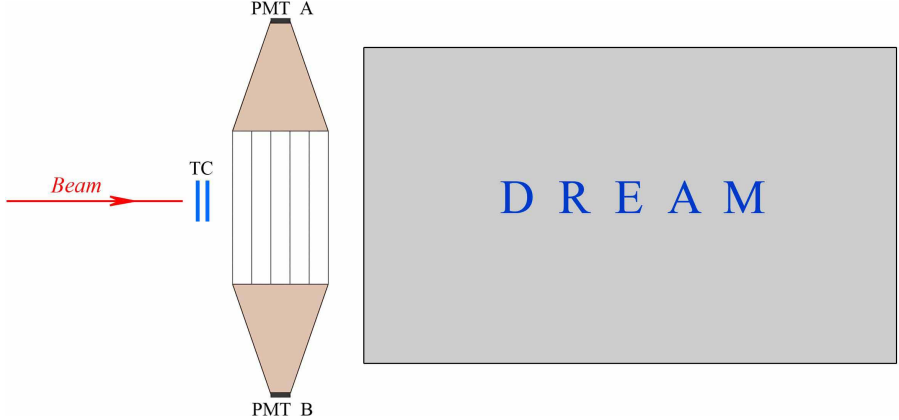}}
\caption{\small
The detector setup used for calibrating the ECAL channels (PMT's A and B).}
\label{setup1}
\end{figure}

The two PMTs reading out the two sides of the ECAL were calibrated with 50 GeV electrons.
To this end, the ECAL was oriented such that the beam entered the detector perpendicular to the 
crystal axis, and the two PMTs collecting the light generated by the showers were located in a plane
perpendicular to the beam axis (Figure \ref{setup1}). 
The main purpose of this setup was to equalize the gain of the two PMTs
in a geometry where the relative contributions of scintillation and \v{C}erenkov light to the signals were the same for both. Given the radiation length of PbWO$_4$ (8.9 mm), the ECAL was only $12.4 X_0$ deep in this geometry, and therefore, a substantial fraction of the shower energy leaked out.
The energy equivalent of the signals thus had to be established on the basis of (EGS4) Monte Carlo simulations.

The 38 PMTs reading out the 19 towers of the HCAL were also all calibrated with 50 GeV
electrons. The showers generated by these particles were not  completely contained in a single calorimeter tower. The
(average) containment was found from EGS4 Monte Carlo simulations. 
When the electrons entered a tower in its geometrical center, on average 
$92.5\%$ of the scintillation light and $93.6\%$ of the \v{C}erenkov light was generated in that tower \cite{DREAMem}.
The remaining fraction of the light was shared by the surrounding towers.
The signals observed in the exposed tower thus corresponded to an energy deposit of 46.3 GeV in the case of the 
scintillating fibers and of 46.8 GeV for the \v{C}erenkov fibers. 
This, together with the precisely measured values of the average signals from the exposed tower, formed the basis for determining the calibration constants, \ie the relationship between the measured number of ADC counts and the corresponding energy deposit.   

\section{Experimental data}

\begin{figure}[htb]
\epsfysize=5.5cm
\centerline{\epsffile{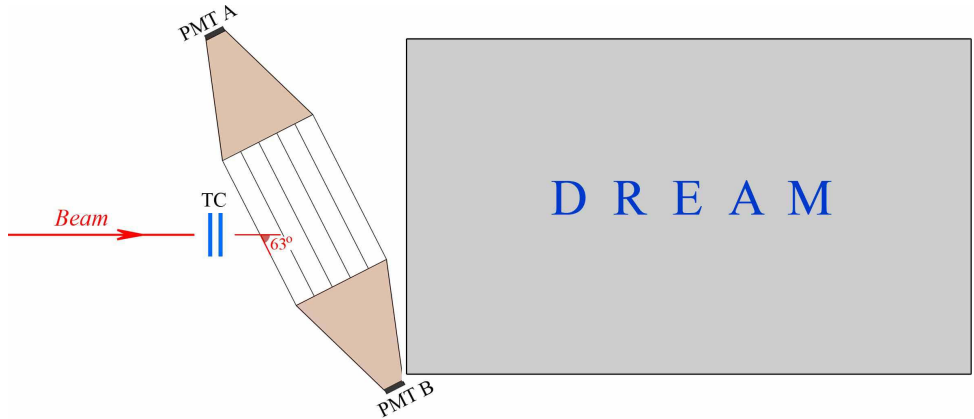}}
\caption{\small
The detector setup used for studying the \v{C}erenkov contributions to the ECAL signals.}
\label{setup2}
\end{figure}

The main purpose of these tests was to see if the ECAL signals could be split into their scintillation and \v{C}erenkov components. In order to optimize the possibilities in this respect, the ECAL was oriented such that the angle between the crystal axis and the beam axis was equal to the \v{C}erenkov angle, 
$\theta_C$: 
\begin{equation}
\cos{\theta_C} = 1/n
\end{equation}
Since the refractive index of PbWO$_4$ is 2.2, $\theta_C \approx 63^\circ$. Detailed measurements
of the angular dependence of the \v{C}erenkov/scintillation ratio \cite{Chproof} revealed indeed a maximum value near $\theta = 63^\circ$, and therefore we chose this angle for our tests.
The detector setup is shown in Figure \ref{setup2}.
In this geometry, the effective depth of the ECAL amounted to $12.4/\sin{\theta_C} \sim 14 X_0$, and 
it contained on average $\sim 90\%$ of 50 GeV electron showers. The remaining 10\% of the shower energy  was recorded in the HCAL.

We exposed the calorimeter system to 50 GeV electrons and to 50 and 100 GeV $\pi^+$. For each run,100 000 events were collected. The time structure of both ECAL signals was measured with 1.25 ns 
resolution. The integrated charge carried by these signals and by those from the 38 HCAL channels was
digitized with 12 bit resolution. 

The beams turned out to be very clean. Contamination in the electron beam was less than 10$^{-4}$.
The pion beams contained muons, at the few-\% level. These muons could be easily recognized and removed from the event samples. We used the muon events for separate studies.

\section{Experimental results}

\subsection{Forward/backward asymmetry}

In lead tungstate, \v{C}erenkov light is emitted at an angle of $63^\circ$ by charged shower particles that are sufficiently relativistic. Unlike the scintillation light, which is isotropically emitted, there is a clear
directional preference in the \v{C}erenkov light production of high-energy showers. Even though much
of the shower energy is deposited by isotropically distributed electrons, produced in Compton scattering
and photoelectric processes, the effects of that on the angular distribution of the emitted \v{C}erenkov light are limited, since much of this takes place below the \v{C}erenkov threshold 
($p_e = 0.26$ MeV/$c$). 
Quantitative information on this was obtained from the angular dependence of the response of quartz-fiber calorimeters (which are exclusively detecting \v{C}erenkov light) to high-energy electron 
showers \cite{Opher}. The maximum response was obtained when the fibers were oriented at the \v{C}erenkov angle with respect to the shower axis. This response was reduced by a factor 2-3 when the fibers were aligned with the beam axis, and by a factor 10 when read out at the upstream end.

We are exploiting this feature by comparing the signals from the two ECAL channels in the $63^\circ$
geometry (Figure \ref{setup2}). Any \v{C}erenkov light produced in the shower development will be preferentially observed
in the downstream channel (to be named channel $B$ in the following), whereas the upstream channel
(channel $A$) will see a much smaller fraction of it. Since both PMTs see the same amount of scintillation light, the {\sl forward/backward asymmetry},
$(B-A)/(B+A)$, is a measure for the fraction of \v{C}erenkov light contained in the signals.
\begin{figure}[htb]
\epsfysize=6.6cm
\centerline{\epsffile{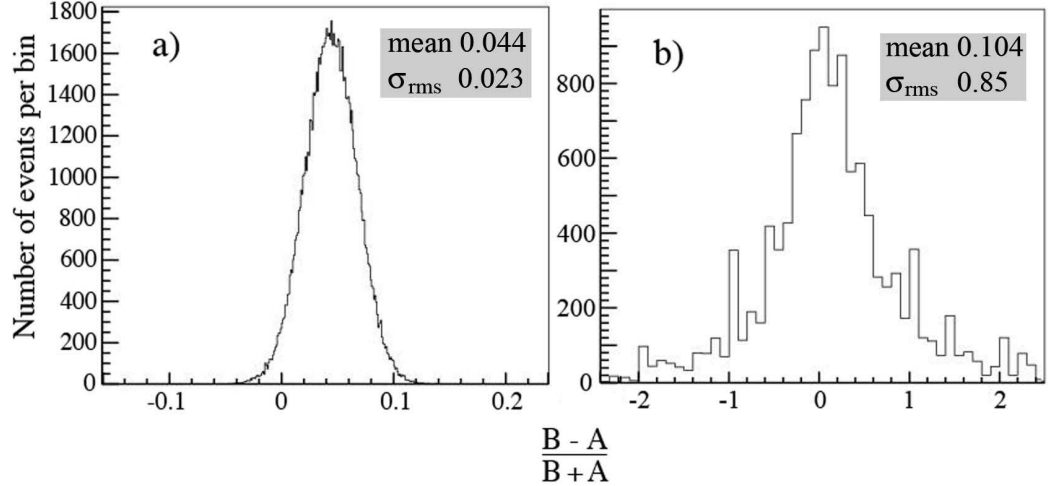}}
\caption{\small
Distribution of the Forward/Backward asymmetry for 50 GeV electron signals ($a$) and 50 GeV $\mu^+$ signals ($b$) in the lead tungstate ECAL.}
\label{Asym_e}
\end{figure}

Figure \ref{Asym_e}a shows the measured asymmetry for the 50 GeV electron signals from the ECAL. Indeed, the $B$ signals are significantly larger than the $A$ ones, on average by $\sim 9\%$.
The forward/backward asymmetry was measured to be $4.4 \pm 0.1\%$.
Forward/backward asymmetry was also observed in the ECAL signals from pions and muons. 
The muon asymmetry is shown in Figure \ref{Asym_e}b. The asymmetry is larger than for electron showers, the $B$ signals were measured to be, on average, larger by $\sim 20\%$ than the $A$ ones
(the forward/backward asymmetry was  measured to be $10.4\pm 1.0\%$). However, the event-to-event fluctuations are considerably larger. The $\sigma_{\rm rms}$ of the distribution for muons is larger by a factor 35-40 than the width of the distribution for electron showers. The latter phenomenon is due to the fact that the electron signals are considerably larger (45 GeV vs 0.1 GeV for a mip), so that Poisson fluctuations in the number of photoelectrons account for at least half of this difference\footnote{A Gaussian fit to the distribution from Figure \ref{Asym_e}b gave a $\sigma$ of 0.60. If (Poisson) photoelectron statistics was the only contributing factor to the width, one should expect a sigma of 0.46.}. In addition, because of the Landau distribution of the muon signals, these statistical fluctuations are non-Poissonian for these particles. 

The larger average asymmetry for muons should be expected on the basis of the fact that \v{C}erenkov light generated in em showers contains an isotropic component,
mainly produced by Compton electrons \cite{RWbook}. In a separate paper, we have shown that measurements of em showers in thin PbWO$_4$ crystals revealed that
the forward/backward asymmetry decreases as the em shower develops. In the first $2-3 X_0$, an asymmetry of $\sim 7\%$ was measured, but this asymmetry decreased considerably when deeper regions of the shower
were probed. The asymmetries measured here for complete em showers and for single particles (muons) are in complete agreement with these observations.
\begin{figure}[htb]
\epsfysize=9.5cm
\centerline{\epsffile{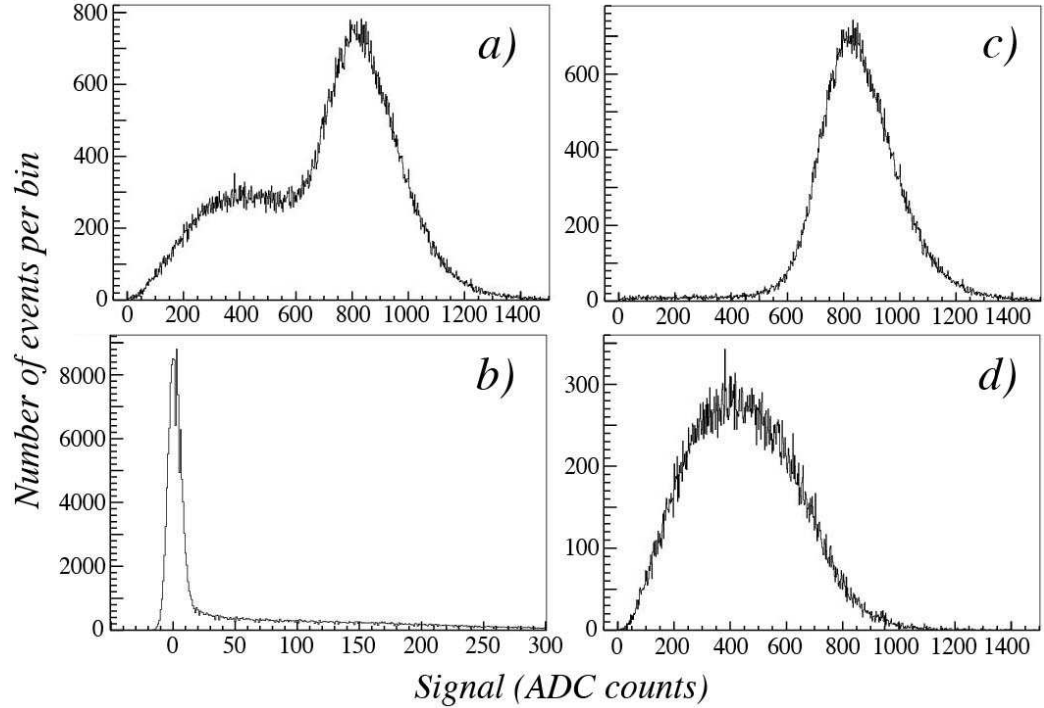}}
\caption{\small
Signal distributions for 50 GeV $\pi^+$ in the ECAL/HCAL system. Shown are the total signal distributions in the HCAL ($a$) and in the ECAL ($b$), as well as the signal distributions in HCAL for
pions that gave a mip signal in the ECAL ($c$) and for pions that produced a larger signal in the ECAL ($d$). The HCAL signals concern the scintillating fibers only. }
\label{pions}
\end{figure}

After removing the (6\%) muons from the 50 GeV hadronic event sample, the remaining pion events exhibited some interesting characteristics, which are a direct consequence of the large $e/h$ ratio
of the PbWO$_4$ crystal calorimeter\footnote{This $e/h$ value was measured to be 2.4 \cite{hcal}.}.
Figures \ref{pions}a and \ref{pions}b show the total signal distributions in the HCAL and ECAL, respectively. The HCAL distributions exhibits a structure that is the sum of two distinct substructures,
as illustrated in Figures \ref{pions}c and \ref{pions}d. Figure \ref{pions}c shows the HCAL distribution
for pions that penetrated the ECAL. These events populate the mip peak in the ECAL distribution. 
The fact that these pions represent 40\% of the total means that the ECAL's thickness in this geometry  corresponded to $-\ln 0.6 = 0.5$ nuclear interaction lengths.

\begin{figure}[htb]
\epsfysize=10cm
\centerline{\epsffile{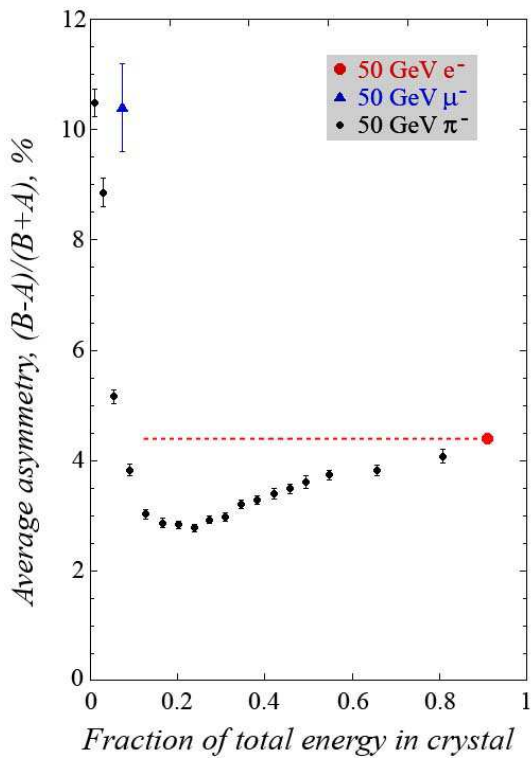}}
\caption{\small
Average Forward/Backward asymmetry for 50 GeV pion signals in the lead tungstate ECAL, as a function of the energy fraction deposited by the beam particles in this detector.}
\label{Asymmetry}
\end{figure}

The pion events were subdivided into samples depending on the total signal produced in the ECAL. 
The underlying idea is that the relatively thin ECAL will, for all practical purposes, only contain the remnants of the first generation of nuclear reactions in the hadronic shower development. If the first interaction took place in the first
two crystals of the ECAL, then $\pi^0$s produced in this first interaction typically deposited most of their energy in the ECAL. Therefore, one should expect a correlation between the total ECAL signal and the fraction of \v{C}erenkov light. Figure \ref{Asymmetry} clearly
exhibits such a correlation. By comparing the results shown in Figures \ref{Asym_e}a and \ref{Asymmetry}, the
average fraction of the ECAL signal carried by em shower components may be determined. 

However, this determination is complicated by another effect shown in Figure \ref{Asymmetry}, namely the Landau tail of 
pions penetrating the ECAL. Not surprisingly, the events in the mip peak of the pion distribution exhibit the same asymmetry as the muons from our sample: $10.5 \pm 0.3\%$, where the smaller error reflects the difference in event statistics.
However, event samples with signals larger than the mip peak in the ECAL do contain some fraction of penetrating
pions, from the Landau tail. Because of the much larger asymmetry for such events, the overall asymmetry for these event samples is measurably affected. A similar effect may be expected for events in which the first pion interaction occurred deep inside the ECAL. Measurements with a single crystal indicated that em
showers developing only over a few radiation lengths exhibit an asymmetry of $\sim 7\%$, \ie  not much smaller than that for mips \cite{Chproof}.

As a result of these complications, the measured asymmetry starts to reflect the em component of the hadron showers only for events in which at least 10 GeV (20\% of the total energy) was deposited in the ECAL.

In order to check the latter statement, we investigated the relationship between the \v{C}erenkov components of the signals observed in the em and hadronic sections of our calorimeter system. In order to limit the contaminating effects of the
penetrating pions, we limited this study to pions that deposited more than 10 GeV in the ECAL. The relative contribution of \v{C}erenkov light to the signals was derived from the $(B-A)/(B+A)$ asymmetry in the ECAL and from the $Q/S$ signal ratio in the DREAM hadronic section. 
\begin{figure}[htb]
\epsfysize=7cm
\centerline{\epsffile{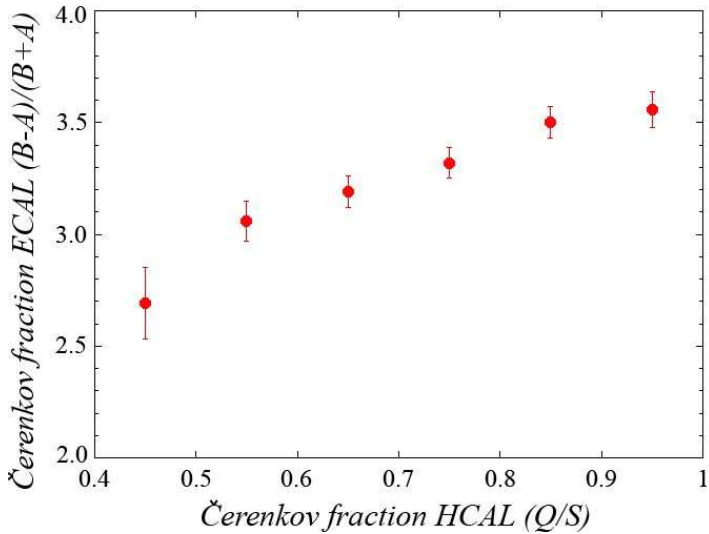}}
\caption{\small
Correlation between the average fractions of \v{C}erenkov light measured in the ECAL and HCAL signals, for 50 GeV pions starting their showers in the ECAL, and depositing at least 10 GeV in the crystals. }
\label{EHCcorr}
\end{figure}
Figure \ref{EHCcorr} shows that there is a clear correlation between these quantities. And since the $Q/S$ signal ratio is directly related to the em shower fraction, $f_{\rm em}$, we conclude from this that the asymmetry measured in the ECAL is indeed an indicator for the fraction of the ECAL signal carried by the em shower component.

\begin{figure}[htb]
\epsfysize=8cm
\centerline{\epsffile{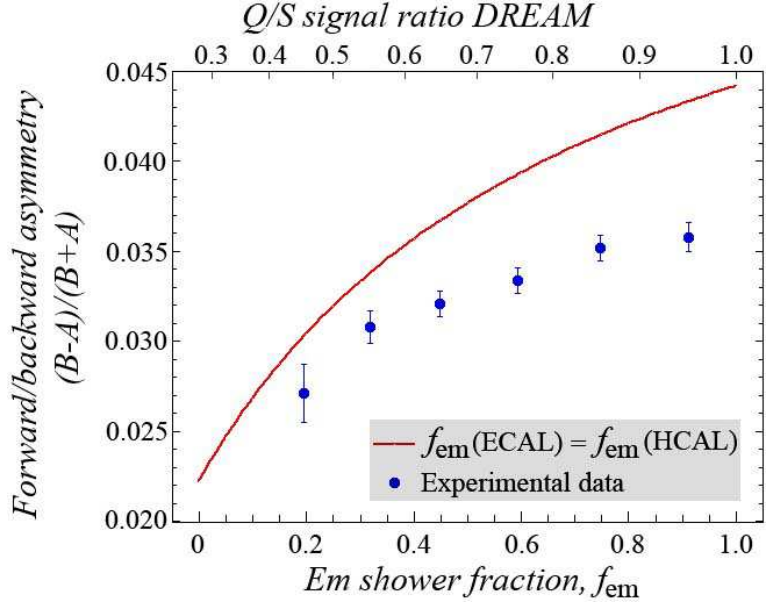}}
\caption{\small
Relationship between measured quantities and the em shower fraction. The measured quantities concern the $Q/S$ signal ratio measured in the DREAM hadronic section (top horizontal axis) and the forward/backward signal asymmetry measured in the crystal ECAL (vertical axis).}
\label{fem_bf}
\end{figure}

This can also be seen as follows. If we denote the ECAL \v{C}erenkov signal by $Q$ and the scintillation signal by $S$, as in the DREAM hadronic section, then the forward/backward asymmetry in the ECAL signals can be written as
\begin{equation}
{{B - A}\over {B + A}} ~=~ {Q\over {2S + Q}}
\label{eq1}
\end{equation}
Since in practice $Q \ll S$, the asymmetry is approximately equal to 0.5 $Q/S$, and therefore the asymmetry is related in a similar way to $f_{\rm em}$ as the $Q/S$ signal ratio measured for the DREAM hadronic section.
In Reference \cite{DREAMhad}, we derived the exact relationship between $Q/S$ and $f_{\rm em}$ for DREAM:
\begin{equation}
{Q\over S} ~=~ {f_{\rm em} + 0.21~(1 - f_{\rm em})\over {f_{\rm em} + 0.76~ (1 - f_{\rm em})}}
\label{eq2}
\end{equation}
where the factors 0.21 and 0.76 are the inverse values of the $e/h$ ratios of the Cu/quartz and Cu/plastic-scintillator
sampling structures, respectively. A similar relationship can be derived between the forward/backward signal asymmetry and the em shower fraction in ECAL. Two modifications of Equation \ref{eq2} are important:
\begin{enumerate}
\item The $e/h$ value of ECAL as a scintillation device is much larger than for the Cu/plastic sampling structure in DREAM: 2.4 \vs 1.3
\item Since the scintillation and \v{C}erenkov signals of the ECAL are not independently calibrated, as in DREAM, an overall
calibration factor is needed that relates the strengths of the $Q$ and $S$ signals in ECAL. This factor is such that
the asymmetry is 0.044 for $f_{\rm em} = 1$ (pure em showers, see Figure \ref{Asym_e}).
\end{enumerate}
These considerations lead to the following relationship between the measured forward/backward asymmetry in the ECAL signals (with the detector oriented at the \v{C}erenkov angle) and the em fraction of the shower energy deposited in this calorimeter section:
\begin{equation}
{{B - A} \over {B + A}} ~=~ 0.044~ {f_{\rm em} + 0.21~(1 - f_{\rm em})\over {f_{\rm em} + 0.42~ (1 - f_{\rm em})}}
\label{eq3}
\end{equation}
This relationship is graphically shown in Figure \ref{fem_bf}. The same figure also shows (on the top horizontal axis) the $Q/S$ ratio in the fiber calorimeter section. 
If the experimental data points were located on this curve, we would conclude that the em shower
fraction derived from the em and hadronic signal characteristics was {\sl exactly} the same. However, since the crystal and the fiber sections of the calorimeter system probed different parts of the shower development, and especially since the crystal section only probed the first 0.5 nuclear interaction lengths, such a perfect one-to-one correspondence should not be expected. However, Figure \ref{fem_bf} does show a clear correlation between the em shower fractions derived from both detector segments.  
%

The practical merits of all this were studied by investigating the relationship between the asymmetry in the ECAL signals
and the {\sl total} response of the calorimeter system (ECAL + HCAL), for this event sample.
We recall that in DREAM, there is a direct relationship between the $Q/S$ signal ratio, \ie the ratio of the total signals measured in the \v{C}erenkov fibers and the scintillation fibers, and the em shower fraction, $f_{\rm em}$ \cite{DREAMhad}. By selecting events with a certain $Q/S$ value, the $f_{\rm em}$ value is fixed. The total calorimeter signal for these events is different from that for events with a different $Q/S$, and thus $f_{\rm em}$ value. 
\begin{figure}[htb]
\epsfysize=9.5cm
\centerline{\epsffile{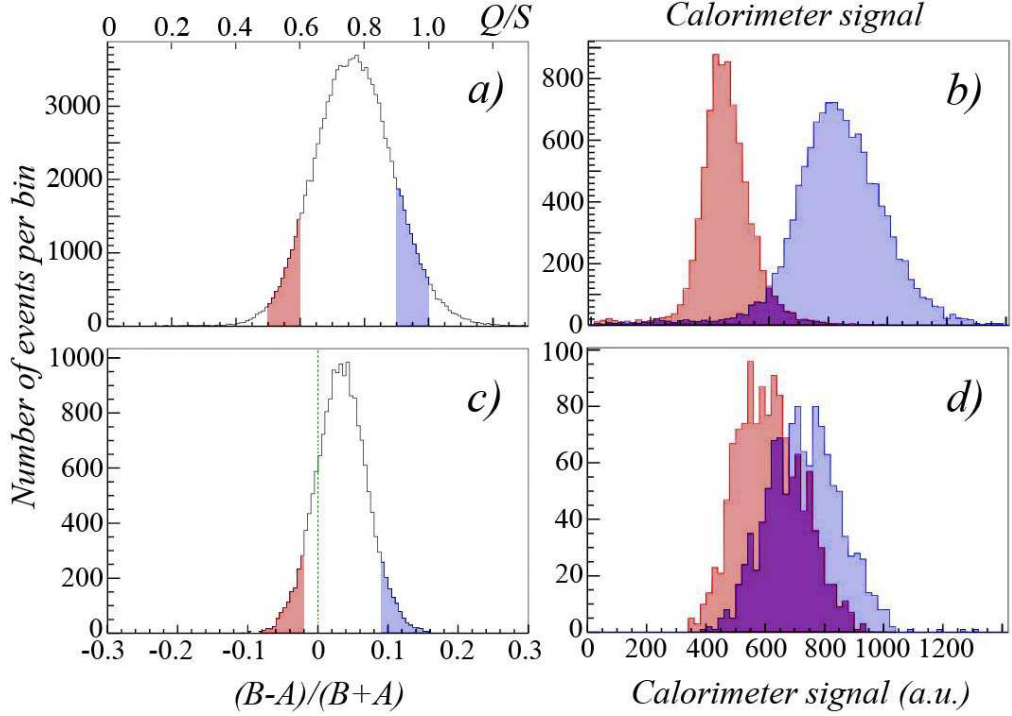}}
\caption{\small
Total calorimeter signal distribution for 50 GeV $\pi^+$ ({\sl right}), for different choices of variables that select the em shower content ({\sl left}). Diagram $b$ shows the total quartz signal for two different $Q/S$ bins, indicated in diagram $a$, for pions that penetrated the crystal ECAL. In diagram $d$, the distribution of the sum of the DREAM scintillator signals and crystal signal $B$ is displayed, for two different choices of the ECAL asymmetry parameter, shown in diagram $c$. The latter distributions concern events in which at least 10 GeV was deposited in the ECAL. }
\label{proof}
\end{figure}

This is illustrated in Figure \ref{proof}, which shows in diagram $b$ two signal distributions for event samples with very different $Q/S$ values. The $Q/S$ bins used for these samples are shown in Figure \ref{proof}a. The events selected for this purpose all concern pions that penetrated the ECAL without a nuclear interaction. Therefore, almost all the energy carried by these pions was deposited in the fiber calorimeter section. These results illustrate that a larger em shower fraction (selected by means of the $Q/S$ signal ratio) leads to a considerably larger total calorimeter signal, especially in the extremely non-compensating copper/quartz-fiber structure ($e/h = 4.7$) used for this purpose.

In Figure \ref{proof}d, two signal distributions are shown for events that were selected on the basis of the
forward/backward asymmetry measured in the ECAL crystal section. The $(B-A)/(B+A)$ bins used for these two samples are shown in Figure \ref{proof}c. Only events in which the pions deposited at least 10 GeV in the
crystal were used for this purpose, and the total signal was calculated as the sum of the signal measured in PMT $B$ and the signal from the scintillating fibers in the hadronic section.
These results exhibit, at least qualitatively, the same characteristics as those shown above for the
penetrating pions, where the $Q/S$ ratio provided information on $f_{\rm em}$: the larger the asymmetry measured in the ECAL signal, \ie the larger the relative contribution of \v{C}erenkov light to the ECAL signals, the larger the em shower fraction, and thus the larger the total calorimeter signal becomes.
\begin{figure}[htb]
\epsfysize=9cm
\centerline{\epsffile{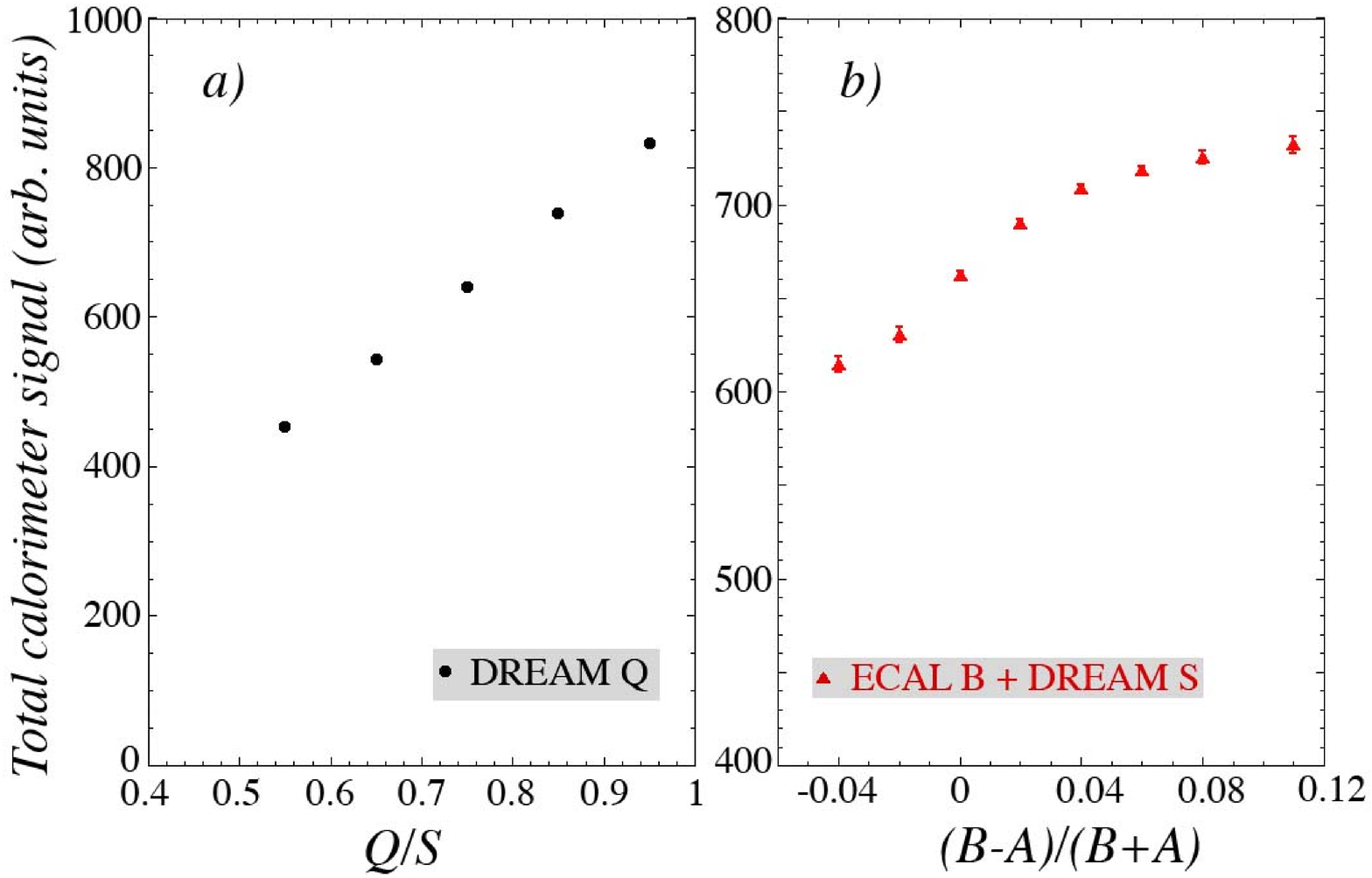}}
\caption{\small
Average total calorimeter signal for 50 GeV $\pi^+$, as a function of variables that select the em shower
content. Diagram $a$ shows the total quartz signal as a function of the $Q/S$ signal ratio, for pions that penetrated the crystal ECAL. In diagram $b$, the sum of the DREAM scintillator signals and the
downstream crystal signal ($B$) is displayed, for events in which at least 10 GeV was deposited in the ECAL. }
\label{DREAMtrick}
\end{figure}

In Figure \ref{DREAMtrick}, the described phenomena are shown for the entire range of possible $f_{\rm em}$ values. The total calorimeter signal for penetrating events (mip in ECAL) is plotted as a function of the $Q/S$ value in Figure \ref{DREAMtrick}a, and in Figure \ref{DREAMtrick}b as a function of the $(B-A)/(B+A)$ value for pions interacting in the ECAL (and depositing more than 10 GeV in the crystals).

\begin{figure}[htb]
\epsfysize=9cm
\centerline{\epsffile{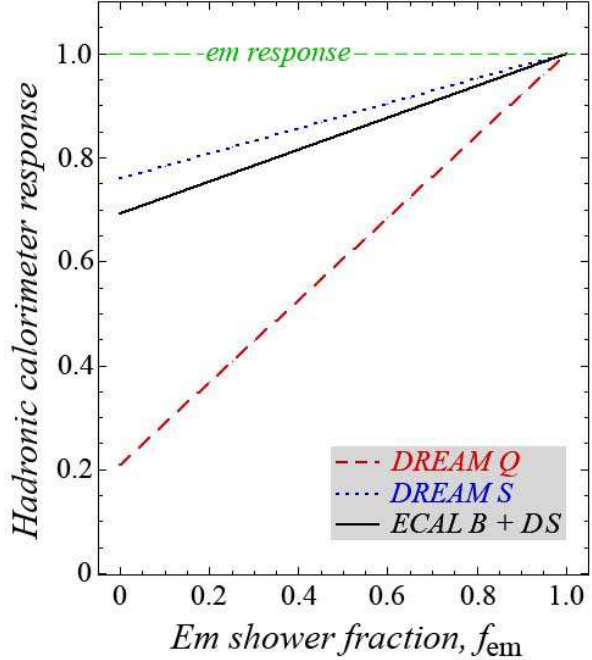}}
\caption{\small
The hadronic calorimeter response as a function of $f_{\rm em}$ for different signals and signal combinations,
which reflect the degree of non-compensation. See text for details.}
\label{Linearity}
\end{figure}

A few remarks are in order. The fact that the effect shown in Figure \ref{DREAMtrick}a (which we will call the {\sl reference effect}) is much larger than that in Figure \ref{DREAMtrick}b can be ascribed to two factors:
\begin{enumerate}
\item The reference effect concerns exclusively \v{C}erenkov light. We are looking at the total \v{C}erenkov signal as a function of $f_{\rm em}$. Had we chosen the signals from the scintillating fibers instead, the observed increase in the calorimeter response would have been only 15\%, instead of the factor of two measured for the \v{C}erenkov reponse \cite{DREAMhad}. This is a consequence of the very large difference between the $e/h$ values (4.7 \vs 1.3), which forms the very basis of the DREAM principle \cite{elba}. 
For the em content derived on the basis of the forward/backward asymmetry in the crystal signals, we don't have the option to look exclusively at \v{C}erenkov light. A very large fraction of the signal from the
crystals consists of scintillation light, even if we optimize the event selection for em shower content, \ie \v{C}erenkov contributions.
\item The ECAL is only $14 X_0$ deep. Even though much of the em component of hadron showers
derives from $\pi^0$ production in the first nuclear interaction of the showering particle, the measurement of $f_{\rm em}$ for the entire shower on the basis of the signals from the first $14 X_0$ has its limitations.
\end{enumerate}
In light of these considerations, the effect observed in Figure \ref{DREAMtrick}b is actually quite remarkable, although it should also be pointed out that, since the $e/h$ value of a homogeneous PbWO$_4$ calorimeter is $\sim 2.4$, a larger increase than the 15\% mentioned above should be expected for the combination considered here. Figure \ref{Linearity} summarizes the non-linearity characteristics for the different signals and signal combinations.

We also want to point to another difference between the data sets shown in Figures \ref{DREAMtrick}a and \ref{DREAMtrick}b. As explained in Reference \cite{DREAMhad}, there is a linear relationship between $f_{\rm em}$ and the total calorimeter signal. Even though the parameters $Q/S$ and\break $(B-A)/(B+A)$ are not exactly proportional to $f_{\rm em}$, it is remarkable that the signal dependence for $Q/S$ is approximately linear, while it is not for $(B-A)/(B+A)$. This may be explained from the fact that the latter curve extends beyond the physically meaningful region, $0 < f_{\rm em} < 1$, contrary to the $Q/S$ curve. This region is limited to $0 < (B-A)/(B+A) < 0.044$, and the points outside this region probe the
tails of the distribution. This may explain the ``S- shaped" curve of Figure \ref{DREAMtrick}b.
A similar deviation from linearity is actually also observed if points covering the unphysical region $Q/S > 1$ are included in Figure \ref{DREAMtrick}a.


\subsection{Time structure of the ECAL signals}

\begin{figure}[htb]
\epsfysize=8cm
\centerline{\epsffile{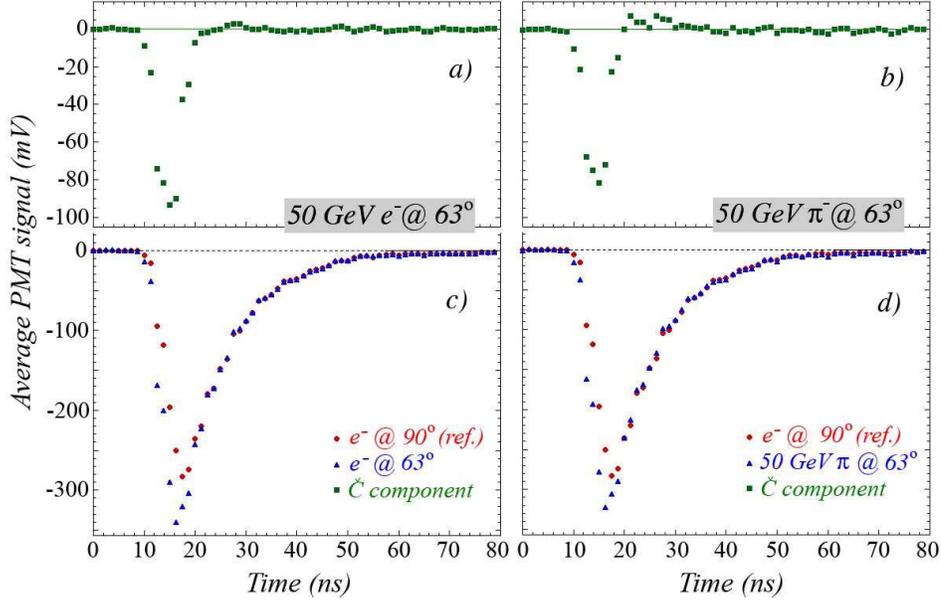}}
\caption{\small
Time structure of the signals from 50 GeV electron showers in the lead tungstate crystals,
measured with detectors located at 90$^\circ$ and 63$^\circ$ with respect to the beam line,
for 50 GeV electron ($c$) and pion ($d$) showers. The top graphs ($a$ and $b$) show the difference between the
time structures recorded at the two different angles, for the electrons and pions, respectively.}
\label{FBtime}
\end{figure}

A second valuable tool for recognizing the contributions of \v{C}erenkov light to the calorimeter signals
is derived from the time structure of the events. This is illustrated in Figure \ref{FBtime}, which
shows the average time structure of the 50 GeV shower signals recorded with the same PMT (B) in the two geometries shown in Figures \ref{setup1} and \ref{setup2}. The two distributions have been 
normalized on the basis of their trailing edge (24-80 ns after the start of the FADC digitization), and are
indeed in great detail identical in that domain, both for the electron (Figure \ref{FBtime}c) and for the pion (Figure \ref{FBtime}d) signals. This part of the pulses is completely determined by the decay characteristics of the scintillation processes in the PbWO$_4$ crystals (see also Figure \ref{esignal})
and should thus be independent of the detector orientation.
However, there is a very significant difference in the {\sl leading} edge of the pulses. The ones measured
in the 63$^\circ$ geometry (Figure \ref{setup2}) exhibit a steeper rise than the ones from the calibration geometry (Figure \ref{setup1}). Figures \ref{FBtime}a and b show the result of subtracting the latter pulse shape from the ``63$^\circ$'' one: the pulses recorded in the 63$^\circ$ geometry contain an additional ``prompt'' component of the type one would expect from
\v{C}erenkov light. In the case of the electron showers, this additional component represents $\sim 11\%$
of the total signal. For comparison, we recall that the forward/backward asymmetry measurements led us to conclude that the signals from PMT B contained, on average,  9\% of \v{C}erenkov light.
The pion signals also exhibit a prompt component, which represents a somewhat smaller fraction of the total signal than in the case of the electrons.
\vskip 1mm
We have studied the possible application of differences in the time structure of the calorimeter signals to determine the
\v{C}erenkov component of the light produced in the crystal calorimeter, in a similar way as we used the measured
measured forward/backward asymmetry (Section 4.1). To this end, we used the time at which the calorimeter signal
reached half of its amplitude value. The larger the relative contribution of \v{C}erenkov light, the earlier this threshold
was reached. Since the contribution of \v{C}erenkov light to the signals from the downstream PMT (B) was larger than that
from the upstream PMT (A), we should thus expect the {\sl difference} between the times at which the two signals reach 50\%
of their amplitude level ($t_A - t_B$), to be a measure for the fraction of \v{C}erenkov light in the signals from the crystal calorimeter.
\begin{figure}[htb]
\epsfysize=7cm
\centerline{\epsffile{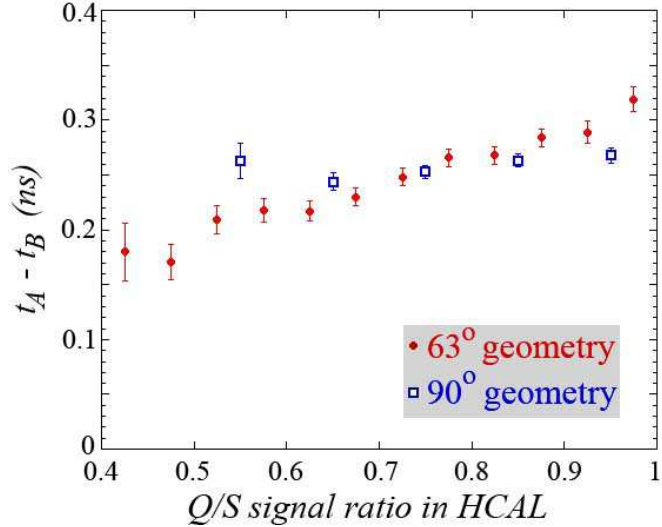}}
\caption{\small
The difference between the times at which the signals from the two PMTs viewing the light produced in the PbWO$_4$ ECAL reach 50\% of their amplitude values, as a function of the fraction of \v{C}erenkov light produced in the showers 
generated by 50 GeV $\pi^+$. This fraction is derived from the $Q/S$ signal ratio in the fiber section of the calorimeter system. Results are given for both geometries used in these studies, described in Figures \ref{setup1} and \ref{setup2}.}
\label{tAtB}
\end{figure}

Figure \ref{tAtB} shows this difference, for 50 GeV $\pi^+$ showers, as a function of the fraction of \v{C}erenkov light,
derived from the $Q/S$ signal ratio measured in the fiber section of the calorimeter. The time difference $t_A - t_B$ increases indeed with this fraction. In order to check the significance of this result, we also repeated this analysis for the measurements done with the same particles (50 GeV $\pi^+$) in the calibration geometry (Figure \ref{setup1}). The results, shown as squares in Figure \ref{tAtB}, indicate no significant dependence on the \v{C}erenkov fraction in the latter (90$^\circ$) setup.

As in the case of the forward/backward asymmetry,
only events depositing at least 10 GeV in the crystal calorimeter section were considered in these analyses, and therefore the results from Figures \ref{tAtB} and \ref{EHCcorr} may be directly compared to each other. 
\begin{figure}[htb]
\epsfysize=8cm
\centerline{\epsffile{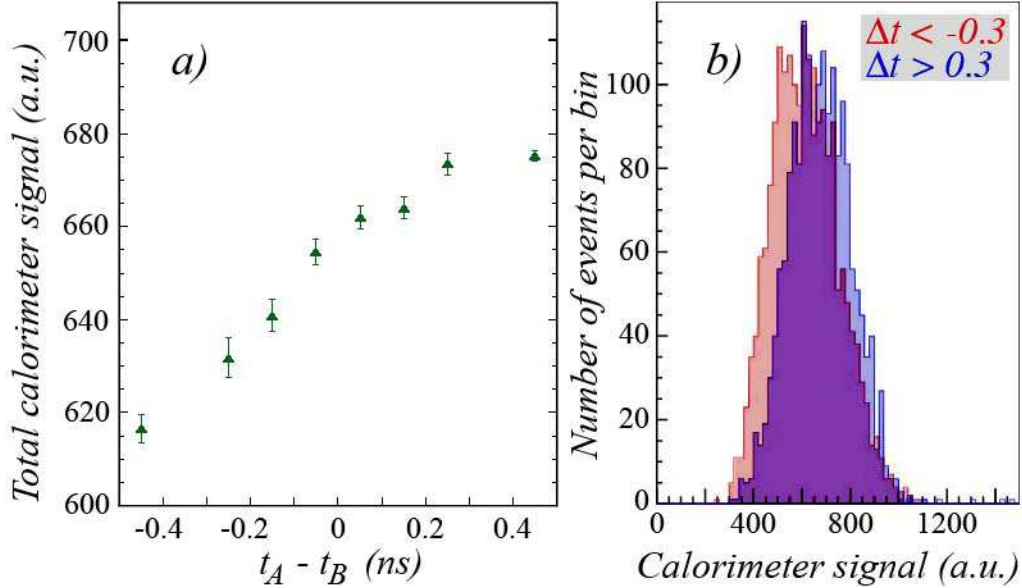}}
\caption{\small
Average total calorimeter signal for 50 GeV $\pi^+$, as a function of the difference between the times at which the
signals from the upstream and downstream PMTs reading out the crystal ECAL reach 50\% of their amplitude values ($a$).
Diagram $b$ shows the total scintillator signal distributions for two subsamples of events, selected on the basis of
this time difference.}
\label{DREAMtrick2}
\end{figure}

We have also studied the correlation between this time difference and the total calorimeter signals. The results are shown in Figure \ref{DREAMtrick2}a, which shows the sum of the signals from ECAL $B$ and the HCAL scintillating fibers, as a function of $t_A - t_B$, for 50 GeV $\pi^+$ showers that deposited at least 10 GeV in the ECAL (early starters). Figure \ref{DREAMtrick2}b shows these total signal distributions for two subsets of events, selected on the basis of the time difference $t_A - t_B$.
These results may be directly compared with those depicted in Figures \ref{DREAMtrick}b and \ref{proof}d, which concern a similar analysis on the basis of the forward/backward asymmetry. We conclude that the time structure of the crystal signals offers equally valuable opportunities for unraveling the crystal signals into their scintillation and \v{C}erenkov components, and thus of an event-by-event measurement of $f_{\rm em}$ as the directionality of the light production.

\section{Conclusions}

We have measured the contribution of \v{C}erenkov light to the signals from electrons, muons and hadrons in an electromagnetic calorimeter made of lead tungstate crystals. In the chosen geometry, which was optimized for detecting this component, information about this contribution was obtained from 
the forward/backward asymmetry in the signals and from their time structure. For single particles traversing the calorimeter (muons, pions), the \v{C}erenkov contribution was measured to be 
$\sim 20\%$. The measurements for showers indicated contributions at about half that level, since
a substantial fraction of the signal is in that case typically  caused by isotropically distributed shower particles.
The pion measurements made it possible to determine the electromagnetic content of the energy deposited in the crystal calorimeter event by event. This information was found to correlate well with 
explicit measurements of this fraction in the dual-readout calorimeter that served as the hadronic section in these measurements. It could also be used in the same way to determine the electromagnetic shower fraction and thus improve the hadronic calorimeter performance. Both the forward/backward asymmetry between the signals from the PMTs reading out both ends of the crystal, and the differences in the time structure of these two signals could be used for this purpose.

\section*{Acknowledgments}

The studies reported in this paper were carried out with PbWO$_4$ crystals made available by the
PHOS group of the ALICE Collaboration. We sincerely thank Drs. Mikhail Ippolitov and Hans Muller 
for their help and generosity in this context.
We thank CERN for making particle beams of excellent quality available.
This study was carried out with financial support of the United States
Department of Energy, under contract DE-FG02-95ER40938.

\bibliographystyle{unsrt}

\begin{thebibliography}{99.}
\bibitem{RWbook} R.~Wigmans, {\em Calorimetry - Energy Measurement in Particle Physics},
International Series of Monographs on Physics, vol. 107, Oxford University Press (2000).

\bibitem{DREAMem} N.~Akchurin \etal, \NIM {\bf A536} (2005) 29.

\bibitem{DREAMhad} N.~Akchurin \etal, \NIM {\bf A537} (2005) 537.

\bibitem{DREAMmu} N.~Akchurin \etal, \NIM {\bf A533} (2004) 305.

\bibitem{Chproof} N.~Akchurin \etal ~(DREAM Collaboration), {\sl Contributions of \v{C}erenkov Light to the Signals from Lead Tungstate Crystals}, submitted to NIM (2007).

\bibitem{Opher} O.~Ganel and R.~Wigmans, \NIM {\bf A365} (1995) 104.

\bibitem{hcal} N.~Achurin \etal, {\sl The Response  of CMS Combined Calorimeters to Single Hadrons, Electrons and Muons}, CMS Internal Note 2007/012 (2007).

\bibitem{elba} R.~Wigmans, \NIM {\bf A572} (2007) 215.











\end{thebibliography}

\end{document}